# ARTICLE

# Pseudo-ternary LiBH$_4$·LiCl·P$_2$S$_5$ system as structurally disordered bulk electrolyte for all-solid-state lithium batteries

Abdel El Kharbachi,*[a,b] Julia Wind,[a,c] Amund Ruud,[c] Astrid B. Høgset,[a] Magnus M. Nygård,[a] Junxian Zhang,[d] Magnus H. Sørby,*[a] Sangryun Kim,[e] Fermin Cuevas,[d] Shin-ichi Orimo,[e,f] Maximilian Fichtner,[b,g] Michel Latroche,[d] Helmer Fjellvåg[c] and Bjørn C. Hauback[a]

The properties of the mixed system LiBH$_4$-LiCl-P$_2$S$_5$ are studied with respect to all-solid-state batteries. The studied material undergoes an amorphization upon heating above 60 °C, accompanied with increased Li$^+$ conductivity beneficial for battery electrolyte applications. The measured ionic conductivity is ~10$^{-3}$ S cm$^{-1}$ at room temperature with an activation energy of 0.40(2) eV after amorphization. Structural analysis and characterization of the material suggest that BH$_4$ groups and PS$_4$ may belong to the same molecular structure, where Cl ions interplay to accommodate the structural unit. Thanks to its conductivity, ductility and electrochemical stability (up to 5 V, Au vs. Li$^+$/Li), this new electrolyte is successfully tested in battery cells operated with a cathode material (layered TiS$_2$, theo. capacity 239 mAh g$^{-1}$) and Li anode resulting in 93% capacity retention (10 cycles) and notable cycling stability under the current density ~12 mA g$^{-1}$ (0.05 C-rate) at 50 °C. Further advanced characterisation by means of *Operando* synchrotron X-ray diffraction in transmission mode contributes explicitly to a better understanding of the (de)lithiation processes of solid-state battery electrodes operated at moderate temperatures.

## Introduction

Substitution of the current liquid electrolytes by solid-state electrolytes (SSEs) is expected to be the next leap in lithium ion battery technology.[1] The integration of SSEs in future batteries is motivated by the expected large energy density, improved safety and stability over a wide temperature range,[2-5] hence exceeding the performances of carbonate or polymer-based (liquid, gel and/or solid) electrolytes.[6-11] The study of complex hydride systems as SSEs has attracted the curiosity of the solid-state ionics community since the discovery of the fast Li-ionic conduction in LiBH$_4$, and *closo*-decaborates.[12-19] A recent work has been reported by Kim et al.[20] for the system, 0.7Li(CB$_9$H$_{10}$)–0.3Li(CB$_{11}$H$_{12}$), showing excellent stability against lithium metal and high conductivity of 6.7 × 10$^{-3}$ S cm$^{-1}$ at 25 °C. This electrolyte system enabled the fabrication of an all-solid-state lithium-sulfur battery with high energy density and cycling stability. Such low-density materials may have a direct impact on the next-generation batteries.

The lithium ionic conduction in the system 90LiBH$_4$-10P$_2$S$_5$ has been studied by Unemoto et al.[21] and fast ionic conductivities were reported for heat-treated samples. The annealing leads to partial decomposition accompanied with loss of B/H species. The addition of LiBH$_4$ or Li(BH$_4$)$_{0.75}$I$_{0.25}$ to Li$_2$S-P$_2$S$_5$ has been shown to increase the ionic conductivities near room temperature (*RT*) and improve the contact at the electrode/electrolyte interfaces during battery tests.[21-23]

The structural, thermodynamic and ionic properties of LiBH$_4$ and its phase transition (orthorhombic to hexagonal, $T_{trs}$ = 113 °C)[24-29] involves a reorientation of the tetrahedral [BH$_4^-$] anions and shortened Li-Li distances with high Li-ionic conduction.[15,30,31] The Li-ion conducting hexagonal phase can be stabilized at *RT* by partly substituting [BH$_4^-$] with the halides Br$^-$ or I$^-$.[32-35] Orthorhombic Li(BH$_4$)$_{1-x}$Cl$_x$, on the other hand, is metastable at *RT* and decomposes to a mixture of LiBH$_4$ and LiCl at a rate which depends on composition. This has been explained by the smaller ionic radius of Cl$^-$ in comparison to I$^-$ and Br$^-$.[36,37] The use of a halide-substituted phase such as Li(BH$_4$)$_{0.75}$I$_{0.25}$ has shown possible solid-electrolyte interface evolving toward the cathode materials during long-term cycling.[38] Although the ionic conductivity of Li(BH$_4$)$_{0.75}$I$_{0.25}$ is several orders of magnitude higher than that of LiBH$_4$,[33,39] the study of composite systems containing the low-*T* modification *ortho*-LiBH$_4$ can be of interest for the design of future batteries,[40] as this latter phase shows pressure-dependent flexible mechanical properties, strain-induced diffusion

[a.] Institute for Energy Technology (IFE), P.O. Box 40, NO-2027 Kjeller, Norway. E-mail: magnus.sorby@ife.no
[b.] Helmholtz Institute Ulm (HIU) Electrochemical Energy Storage, Helmholtzstr. 11, 89081 Ulm, Germany. E-mail: kharbachi@kit.edu
[c.] Centre for Materials Science and Nanotechnology, University of Oslo, P.O. Box 1126, Blindern, NO-0318 Oslo, Norway.
[d.] Univ Paris Est Creteil, CNRS, ICMPE, UMR7182, F-94320, Thiais, France.
[e.] Institute for Materials Research, Tohoku University, Sendai 980-8577, Japan.
[f.] WPI-Advanced Institute for Materials Research, Tohoku University, Sendai 980-8577, Japan.
[g.] Institute of Nanotechnology, Karlsruhe Institute of Technology (KIT), P.O. Box 3640, 76021 Karlsruhe, Germany.

activation energy and formation of a stable interface with many promising electrodes.[41-48]

Here, we report our recent findings regarding the ionic conductivity of the pseudo-ternary system $LiBH_4$-$LiCl$-$P_2S_5$ for solid-state battery electrolytes. The prepared electrolytes are investigated with respect to their structural and ionic properties and cycling stability in lithium metal cells. Thanks to the low scattering of the bulk electrolyte, a pioneering *operando* synchrotron X-ray diffraction study is presented to obtain a better understanding of the processes in the assembled solid-state batteries under operation at moderate temperatures.

## Experimental

### Materials synthesis and characterization

$LiBH_4$ (95%), $LiCl$ (99.9%) and $P_2S_5$ (99%) were purchased from Sigma-Aldrich and stored in an Ar-filled glove box (<1 ppm $O_2$, $H_2O$). $LiBH_4$ and $LiCl$ in 3:1 molar ratio were ball-milled for 5 hours using a Fritsch Pulverisette 6 (P6) planetary ball-mill with stainless steel vials and balls (ball-to-powder ratio 40:1, 370 rpm). Crystalline $P_2S_5$ was then added to this mixture and ball-milled with Spex-mill for 3 hours with the same conditions. All the preparations were carried out in the glove box. Two compositions (mole ratios) were prepared:
80(3$LiBH_4$·$LiCl$)·20$P_2S_5$: called LCPS20
90(3$LiBH_4$·$LiCl$)·10$P_2S_5$: called LCPS10

Synchrotron radiation powder X-ray diffraction (SR-PXD) patterns were obtained at the Swiss-Norwegian Beamlines (SNBL, BM31), ESRF, Grenoble, France with a Dexela 2-dimensional CMOS detector,[49] and a wavelength of 0.3123 or 0.4943 Å calibrated against a NIST Si standard. The samples were contained in 0.5 mm boronglass capillaries that were rotated 90 degrees during the 30 second exposure. The sample-detector distance was 345.97 mm. 1D data were obtained by integration of the 2D diffraction patterns with the program Bubble.[50] Phase identification from the PXD data was performed using the DIFFRAC.SUITE EVA software with the PDF-4 database.

*Operando* SR-PXD data were collected at the SNBL BM01B. All-solid-state batteries were assembled into a dedicated electrochemical cell of Swagelok-type with Kapton windows.[51] Cells were heated to a temperature of 60 °C (in-house built heater) and allowed to equilibrate for 60 minutes. Cycling was performed using a Bat-Small battery cycler (Astrol) with an applied C-rate current of C/10. All profile fittings and Rietveld refinements were performed within TOPAS V5 (Bruker AXS). For SR-PXD patterns, background (Chebychev polynomial), zero-shift, peak-profile parameters, unit cell parameters and scale factor were refined. Broad background features (due to amorphous contributions) were fitted with Gaussian peaks (refined position and broadening). During sequential refinement across all collected operando patterns, only unit cell dimensions and scale factors were refined. All the samples for PXD were assembled and sealed under Ar in the glove box.

TGA/DSC thermal analysis was performed with a Netzsch STA 449 F3 Jupiter instrument in the 25–400 °C temperature range. Samples were measured with Al crucibles at a heating rate of 10 °C min$^{-1}$ and 50 mL min$^{-1}$ Ar flow. Characterizations of the vibrational states were performed by Raman spectroscopy (Nicolet Almega-HD, Thermo Scientific) using a dedicated cell without any air exposure.

### Electrochemical analysis and battery tests

Ionic conductivities were determined by electrochemical impedance spectroscopy (EIS). The powder samples were pressed into 8 mm diameter pellet of < 2 mm thickness by a uniaxial press at around 240 MPa inside the glove box at ambient temperature. The pellets were sandwiched by lithium foils as non-blocking electrodes and sealed in a homemade cell without air contact.[52] The cells were placed in the heating jacket and the EIS were carried out over a frequency range from 1 MHz to 4 Hz using a HIOKI 3532-80 from *RT* to 150 °C in heating and cooling runs. A program-interface allows the automatic control of the stability of each measurement at fixed temperature in equilibrium conditions.

The measured impedance spectra were analyzed by equivalent circuits using the ZView2 software (Scribner Associates Inc.). Additional EIS measurements and cyclic voltammetry were performed using Bio-Logic® VSP multi-channel potentiostat, either in coin cells or a homemade cell described elsewhere.[53]

For battery tests, $TiS_2$ (99.9%, Sigma-Aldrich) and Li foil were used as working and counter/reference electrode, respectively. The $TiS_2$ and the prepared SE powders in 2:3 mass ratio were hand mixed in an agate mortar inside the glove box. The obtained mixture was used as the electrode composite. Around 6 mg of this composite and 30 mg of the SE were introduced in a 10 mm die set and uniaxially pressed together at 240 MPa. A Li foil was placed on the opposite side of the electrode composite before the pellet was inserted in a coin cell. The assembled cells were moved out of glovebox and annealed at 60 °C for 5 h before testing at a cycling station from Bio-Logic Instrument in a temperature-controlled cabinet at 50 °C.

## Results and discussion

Fig. 1 presents the SR-PXD patterns of ball milled 3$LiBH_4$-$LiCl$ and the prepared LCPS10 sample. For the 3$LiBH_4$-$LiCl$ sample the pattern shows Bragg peaks from $LiBH_4$ and $LiCl$ without any noticeable formation of the $Li(BH_4)_{1-x}Cl_x$ phase (Fig. 1a), in agreement with previous studies of the $LiBH_4$-$LiCl$ pseudo-binary system.[36,37,52,54] The phase composition estimated from Rietveld refinements is 65 mole% of $LiBH_4$ and 35 mole% of $LiCl$. This difference compared to the nominal composition before ball-milling could be attributed to the presence of an amorphous part and/or a metastable phase at low content. No additional peaks are present. On the other hand, the addition of $P_2S_5$ to this system and further ball-milling (sample LCPS10) leads to almost complete disappearance of $LiBH_4$ peaks (Fig.1b) and increased diffuse scattering from an amorphous phase. However, the peaks from the $LiCl$ phase remain intense. In

Journal Name | ARTICLE

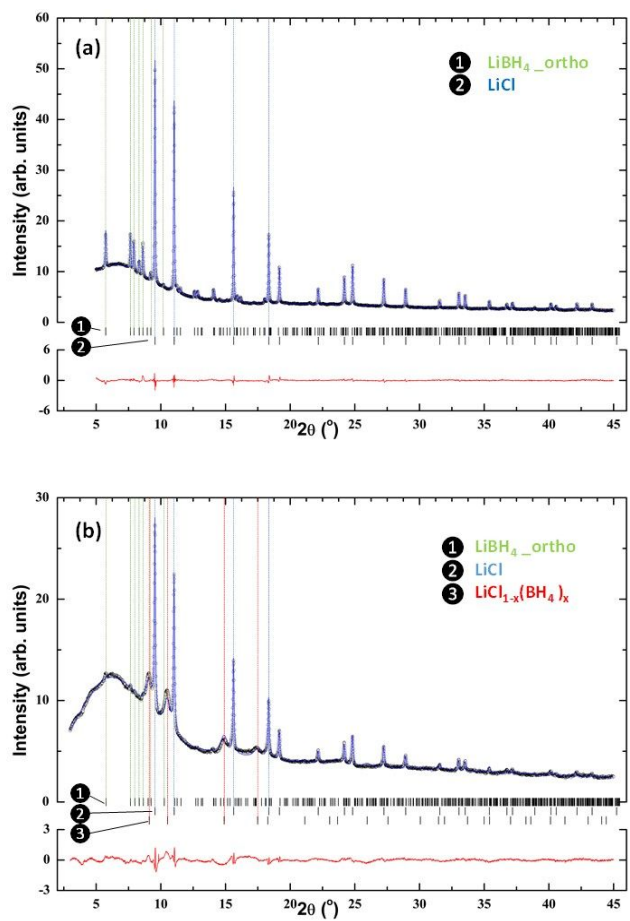

Figure 1. SR-PXD patterns of (a) 3LiBH$_4$-LiCl as mixed after ball-milling and (b) LCPS10 solid electrolyte. The collected data at *RT* are shown in black, the fitted curve in blue, peak markers are shown in order for LiBH$_4$, LiCl and LiBH$_4$/LiCl, respectively. The difference plot (obs-calc) is shown in red at the bottom (λ = 0.4943 Å).

addition, a set of broader peaks, most prominent at lower angles, is observed. Being located at slightly lower angles compared to the LiCl phase, these can be attributed to a LiCl-like structure with larger unit cell dimensions, such as a solid solution of LiCl$_{1-x}$(BH$_4$)$_x$, with or without PS$_4$ anions.

Previous studies have established a Vegards' law behavior for LiCl$_{1-x}$(BH$_4$)$_x$,[55,56] and the unit cell parameter of $a$ = 5.385(2) Å for the present phase (compared to 5.1419(1) Å for pure LiCl) indicates a composition of LiCl$_{0.6}$(BH$_4$)$_{0.4}$. Based on this assumption, the estimated crystalline phase fractions in sample LCPS10 are 26.8 mole% LiCl, 66.7 mole% LiCl$_{0.6}$(BH$_4$)$_{0.4}$ and 6.5 mole% LiBH$_4$. Possible structural trends for compositions with varied P$_2$S$_5$ contents (5-50 mol.%) have been investigated (Fig. S1 in SI section). For LCPS20 the collected PXD pattern is dominated by diffuse scattering from an amorphous phase with minor Bragg peaks of LiCl. At higher P$_2$S$_5$ contents, only diffuse scattering is observed.

The ionic conductivities of both LCPS10 and LCPS20 were measured in the temperature range *RT*-150°C (Fig. 2a/b). The two samples show comparable conductivity at *RT*, in the order of 10$^{-5}$ S cm$^{-1}$. While the conductivity for LCPS20 does not change during the first 2 cycles, the conductivity for LCPS10 increases after the 1$^{st}$ heating and becomes stable for the next cooling/heating/cooling sequences.

For LCPS20, the measurements show stable conductivity during temperature cycling with almost no hysteresis. The LCPS10 SSE presents a slight hysteresis in the second heating/cooling run. With exception of the first heating on LCPS10, the conductivity plots versus 1/*T* are linear, thus indicating that no phase transition between orthorhombic and hexagonal LiBH$_4$ is taking place. The assumption of the elimination of the phase transition in LiBH$_4$ by anion substitution agrees with previously reported studies on halide substitution.[32,57] Worth mentioning that the resulted powders are whitish for LCPS10 and brownish for LCPS20. We suspect the formation of a new dominant amorphous phase in LCPS20 during synthesis (Fig. S1 in SI section), which is stable during heating. In contrast, structural changes for LCPS10 occur during heat treatment. Although the conductivity range for LCPS20 tends to fall in the same order of magnitude as for the reported one for the Argyrodite Li$_6$PS$_5$Cl$_{0.83}$(BH$_4$)$_{0.17}$ with low BH$_4$/Li ratio,[58] however owing to the differences in composition, structure and materials processing and synthesis, the comparison becomes tricky.

Furthermore, this ionic behavior reflects the structural difference between the two SSE materials. For LCPS10, the final conductivities are significantly higher than the reported values for LiBH$_4$ and LiBH$_4$/LiCl pseudo-binary system at the same temperatures.[22,52,59] Correspondingly, the obtained conductivities follow an Arrhenius trend (Fig. 2c) according to the relation: $\sigma T = \sigma_0 \exp\left(-\frac{E_a}{kT}\right)$, where σ is the ionic

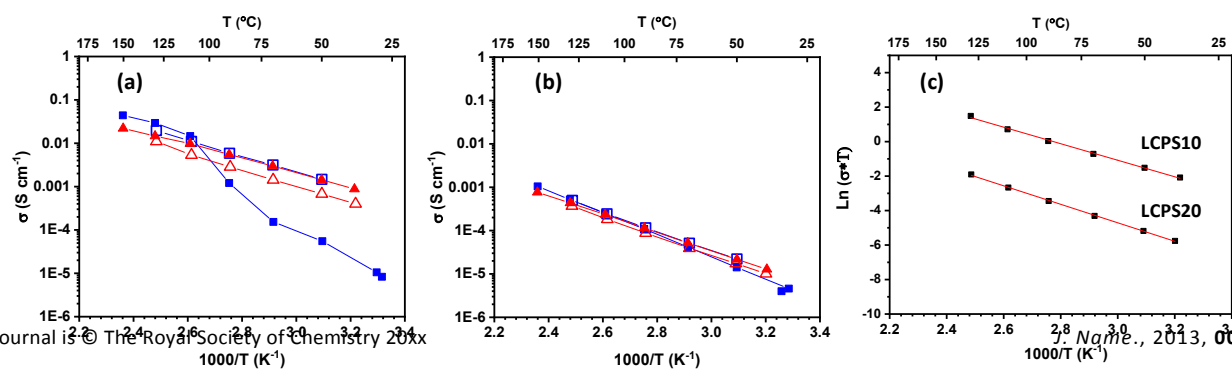

Figure 2. Temperature dependence of the ionic conductivities for (a) LCPS10 and (b) LCPS20 during the first two heating/cooling cycling T ramps. Filled/open squares and triangles correspond to heating/cooling for the 1$^{st}$ (blue) and 2$^{nd}$ (red) runs, respectively. (c) Arrhenius plots of the 2nd cooling used to infer the activation energies (E$_a$) for LCPS10 and LCPS20 samples.

ARTICLE                                                                                                  Journal Name

conductivity and $E_a$ the activation energy. The calculated $E_a$ for LCPS10 and LCPS20 SSEs are 0.40 (2) eV and 0.49 (2) eV respectively. Overall, a remarkable ionic conductivity of about $10^{-3}$ S cm$^{-1}$ near *RT* is reached for LCPS10. This result may have a direct implication on the development of a new generation of solid-state batteries with high-energy density and a wide temperature range which cannot be achieved by current commercial carbonate-based liquid electrolytes.[60,61] Based on this fact, the following investigations will focus on the LCPS10 sample.

TGA/DSC thermal analysis (Fig. 3) was carried out for LCPS10 samples, both as-milled and after annealing at 150 °C for 8 hours under $H_2$ atmosphere in a closed vessel. The ball-milled sample displays two small events at 60 °C (endothermic) and 105 °C (exothermic), the second being accompanied with 2.5% mass loss. However, the pre-annealed sample at 150 °C is stable in this temperature range and showing a mass loss only above 200 °C. Additional exothermic event can be seen at 280-300 °C.

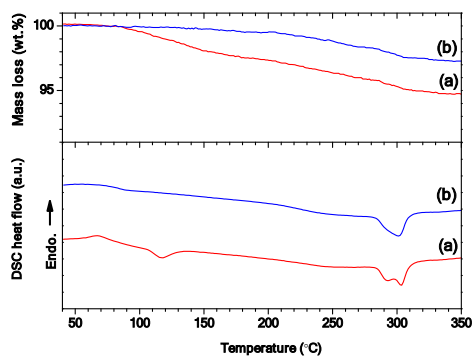

Figure 3. TGA (top) and DSC (bottom) analysis for (a) ball-milled LCPS10 and (b) annealed LCPS10 at 150 °C.

The SR-PXD patterns of LCPS10 collected during heating from *RT* to 150 °C with a rate of 10 °C min$^{-1}$ are shown in Fig. 4. At *RT* before heating, the sample shows Bragg peaks from LiCl, a solid solution-like phase LiCl$_{1-x}$(BH$_4$)$_x$ (Fig. 1b) and pronounced diffuse scattering from amorphous components. During heating of the sample, a gradual structural transformation is observed, in excellent agreement with conductivity and DSC measurements. LiCl Bragg peaks disappear and the sample undergoes total amorphization at around 80 °C, which seems not related to mass loss according to TGA. Based on DSC and PXD analysis, it seems the high-*T* LiCl$_{1-x}$(BH$_4$)$_x$ phase with higher free energy is playing a key role in inducing this transformation in the presence of 10 mole % $P_2S_5$ and facilitates thermodynamically the formation of the final material. Excess $P_2S_5$ would lead to a Li-poor phase with lower conductivities (Fig. S1 and Fig. 2b). Besides the observed irreversibility of the process upon cooling down to *RT*, the TGA/DSC analysis does not show any significant thermal event for the annealed LCPS10 in the temperature range *RT*-150 °C, indicating a good thermal stability for high-*T* applications.

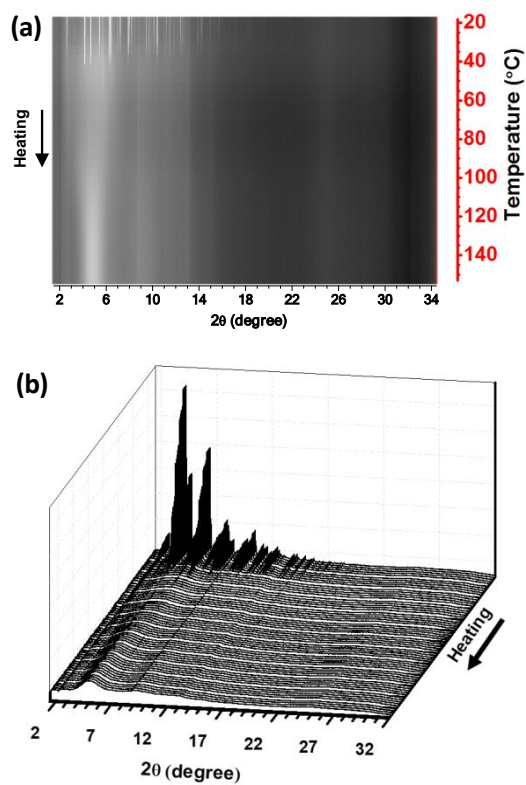

Figure 4. 2D (a) and 3D (b) views of the SR-PXD of the LCPS10 as function of temperature from RT to 150 °C (10 °C min$^{-1}$; λ =0.3123 Å). The linear heating scale is shown on the right Y-axis in (a).

Raman spectroscopy has been carried out to elucidate the chemical environment of the BH$_4$, PS$_4$ and/or $P_2S_6$ groups in the LCPS10 and LCPS20 samples. The obtained spectra, compared to those of the starting materials 3LiBH$_4$-LiCl and $P_2S_5$, are shown in Fig. 5. 3LiBH$_4$-LiCl shows the same [BH$_4^-$] vibrational modes as pure LiBH$_4$,[22,23] but with broader and less well-defined bands in accordance with previous works.[34,35] The two characteristic bands (stretching and bending) of pure LiBH$_4$ can be seen in the 3000-400 cm$^{-1}$ spectral region. The stretching band (1350-1000 cm$^{-1}$) is split across a wide region owing to the presence of an overtone (3$v_L$, 1231 cm$^{-1}$).[62,63] The mixing with $P_2S_5$ at different proportions leads to weaker intensities and disappearance of some features of the [BH$_4^-$] vibrational modes. However, the two main bands are still present, and the lowered intensities may suggest a decrease of the symmetry and a modified [BH$_4^-$] geometry. The most reported [PS$_4^{3-}$] entity is characterized by a wide peak at 542 cm$^{-1}$ and is clearly observed in the spectrum of the LCPS10 sample.[64]

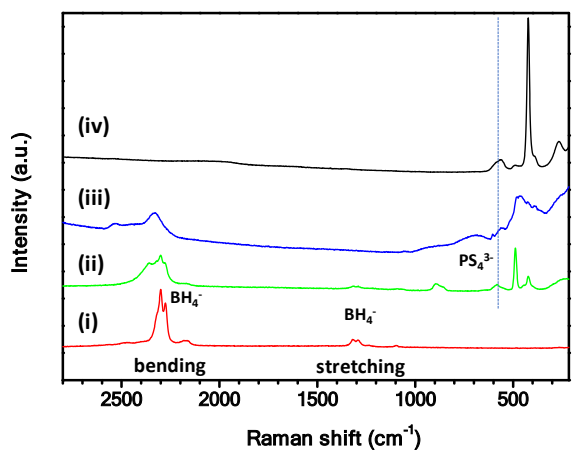

Figure 5. Raman spectra region 2800-210 cm$^{-1}$ of (i) 3LiBH$_4$-LiCl, (ii) LCPS10, (iii) LCPS20 and (iv) P$_2$S$_5$ samples.

higher structural disorder (amorphisation) for LCPS20 in agreement with the PXD analysis (Fig. S1). The peak around 500 cm$^{-1}$ can be linked to the S-S bond and the shoulder at 2385 cm$^{-1}$ in LCPS10 is attributable to a feature of the stretching band which agree with the fact that [BH$_4^-$] ions may exist in different chemical environments,[22] likely formed from the incorporation of [PS$_4^{3-}$] ions. Based on the observed DSC event described above in the *RT*-150 °C temperature range (Fig. 3), the observed irreversible transformation in the SR-PXD suggests a subtle rearrangement of the BH$_4$ groups of the high-*T* hexagonal LiCl$_{1-x}$(BH$_4$)$_x$ phase in the presence of [PS$_4^{3-}$] anions.[37] This assumption is corroborated by Raman spectroscopy where the BH$_4$ and PS$_4$ groups may co-exist in close chemical proximity *i.e.* the same phase. However, it is not possible from the present diffraction and Raman data to establish the crystal structure of the new formed phase(s) in LCPS10 electrolyte.

The peak attributed to the [PS$_4^{3-}$] group in LCPS20 becomes broad and slightly shifted. This agrees with higher configurational disorder. The [P$_2$S$_6^{4-}$]-anions may be present as indicated by the band at 385 cm$^{-1}$. However, at low wavenumbers, Raman spectra confirm explicitly the presence of [PS$_4^{3-}$] moieties represented by the three peaks at 267, 422 and 564 cm$^{-1}$ in the LCPS10 sample,[16,59,64] as well as an indication of

Further electrochemical characterizations were focused on the sample LCPS10 in order to obtain more details on the behaviour in the battery cell. The study is supplemented by electrochemical stability measurements and battery tests at 50 °C. Prior to being used in a battery cell, the SE powders were homogenized by heat treatment at 150 °C for 24 h in reducing atmosphere (2 MPa H$_2$). Using this optimized SSE, battery tests were carried out in a two-electrode Li-ion cells using TiS$_2$ cathode and Li anode (CE/Ref). TiS$_2$ adopts a trigonal layered structure (space group $P\text{-}3m1$, No. 164) with the initial lattice parameters $a = b$ = 3.4073 and $c$ = 5.6953. Up to one Li can be inserted assuming the complete reaction: $yLi + TiS_2 \rightarrow Li_yTiS_2$.[65] Having good electronic conductivity,[66] this electrode can be used without further carbon additions; hence reducing the number of interface issues that can be encountered in solid-state batteries.[67,68] Fig. 6a/b shows the galvanostatic charge/discharge cycling at rate C/20 (10 cycles) for two cells with thin (300 µm) and thick (600 µm) bulk SSE. Both cells show remarkable cyclability and stability over several cycles. Note that the electrolyte presents some ductility, such a property could be suitable to compensate the electrode volume changes as a major issue of solid-state batteries.

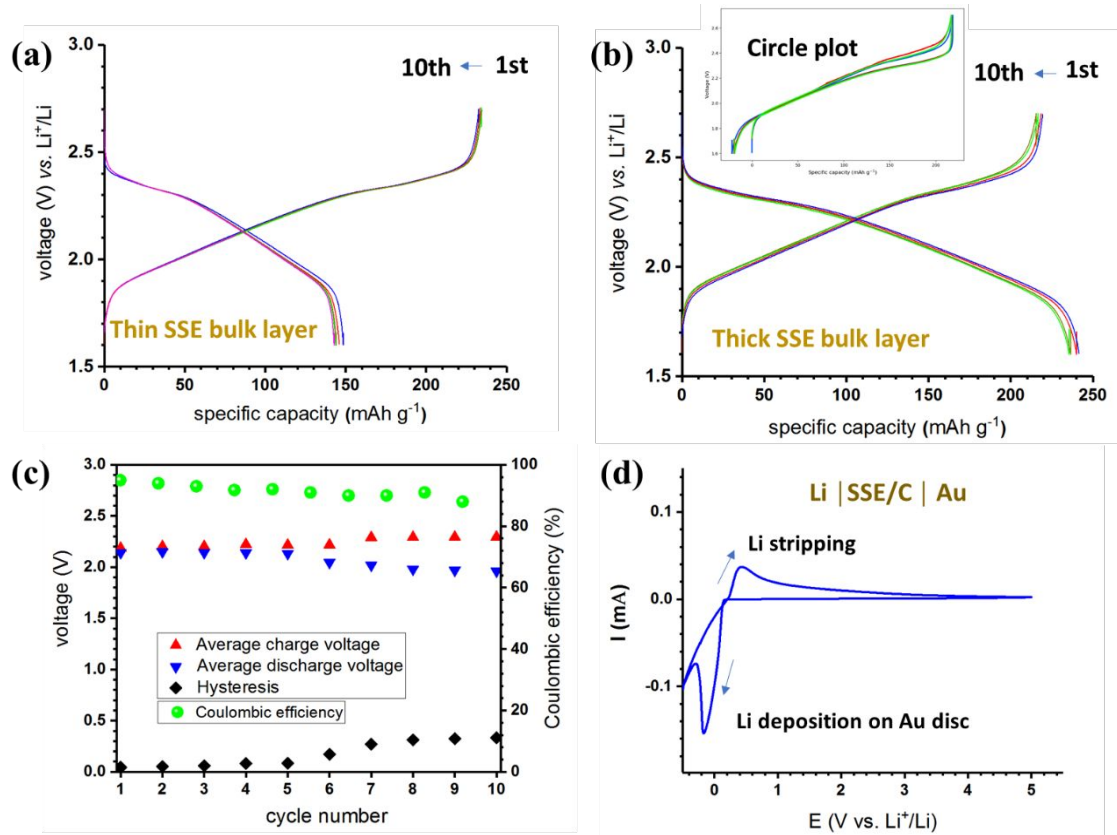

Figure 6. Electrochemical characterization and performance of the LCPS10 SSE: Galvanostatic discharge/charge cycling (0.05 C-rate) at 50 °C with TiS$_2$ electrode (a) using thin (300 μm) and (b) thicker (600 μm) bulk layer of SSE; For the later cell, (c) average voltage, hysteresis (ΔV) and Coulombic efficiency as function of cycle number; (d) cyclic voltammetry at Au disc as working electrode *vs.* Li$^+$/Li.

The first cell (Fig. 6a) shows lower capacity during discharge compared to the charge process. Since this seems to be stable and reproducible, one could attribute such behavior to the increase of the stress in the working electrode during lithiation as the cells were pressed at the same pressure. In fact, thin bulk cells may allow cycling at higher rates, but for developments and applications the engineering of the cell configuration need to be optimized. For the second cell (Fig. 6b/c), the capacity retention presents 93% of the theoretical capacity of TiS$_2$ (~239 mAh g$^{-1}$) after 10 cycles, while the Coulombic efficiency remains stable during cycling. Based on the average voltages, the discharge-charge hysteresis is significantly smaller than usual. Cyclic voltammogram of the SSE is presented in Fig. 6d, and the stripping/platting of Li can be observed around 0 V with successive oxidation and reduction of the Li metal.

A layer of mixed SSE-carbon black has been added to ensure better adhesion to the Au surface. At low voltage, a possible side reaction may be due to the presence of carbon and/or impurities. However, at high voltage, the "active material-free" LCPS10 electrolyte shows an electrochemical window up to 5 V.

Fig. 7 shows *operando* SR-PXD patterns of the battery cell during the discharge presented in Fig. 6b. The sketch of the experimental set-up is shown in Fig. 7a. The first diffraction pattern (before cycling starts) indicates that the main contributions are LiCl and TiS$_2$ in addition to the current collector (Al). Significant shifts in the Bragg peaks, corresponding to expansion of the TiS$_2$ electrode (Li$_y$TiS$_2$) along the $c$-axis during lithiation can be observed. Miller indices of the main reflections as well as arrows indicating the reflections with the main changes are shown in Fig. 7b.

The corresponding change in the lattice parameter $c$ as a function of the discharge capacity is shown in Fig. 7c. In fact, the TiS$_2$ expansion occurs only in $c$ direction with no changes in $a$ direction, due to the layered structure of the material. As expected, no changes in the bulk SSE itself can be observed – the SSE peaks remain stable throughout the *operando* experiment. With the present setup, no additional potential side reactions and/or interface formations and evolutions can be detected. Owing to beam time constraints, the cells were only lithiated partially.

Data for the re-charge of the cell, showing the corresponding contraction along the $c$-axis for Li$_y$TiS$_2$ during the delithiation process, are summarized in Fig. S2. Accordingly, one can witness a change in the expansion along the $c$-axis for the Li$_y$TiS$_2$ solid solution but no change in the bulk electrolyte material itself. The observed instabilities for the $c$ parameters during delithiation (Fig. S2b), may be attributed to inhomogeneities and segregations, which could reflect changes in the intrinsic properties between the charged and discharged states.

## Conclusions

A pseudo-ternary LiBH$_4$-LiCl-P$_2$S$_5$ solid-state electrolyte is reported in this study. The as-milled materials showed low ionic conductivity. However, a high ionic conductivity of ~10$^{-3}$ S cm$^{-1}$ with an activation energy of 0.40(2) eV was measured at RT after heat treatment at 150 °C. This solid-state electrolyte has the approximate nominal composition $(LiBH_4)_{0.73} \cdot (LiCl)_{0.24} \cdot (P_2S_5)_{0.03}$ and usable conductivity range for applications in solid state batteries. Structural analysis and characterization of this material suggest that $BH_4$ and $PS_4$ groups may belong to the same structural unit, while Cl ions substitution by $PS_4$ may occur in this structure when small amounts of $P_2S_5$ are added. During the preliminary electrochemical tests, this new SSE is stable in contact with Li metal and battery tests using $TiS_2$ cathode show notable cyclability and reversibility. Furthermore, thanks to the low scattering of the bulk electrolyte based on light-weight borohydride, we demonstrate the possibility of *operando* PXD analysis in transmission mode of a solid-state battery, where (de)lithiation processes can be followed during the battery cycling tests.

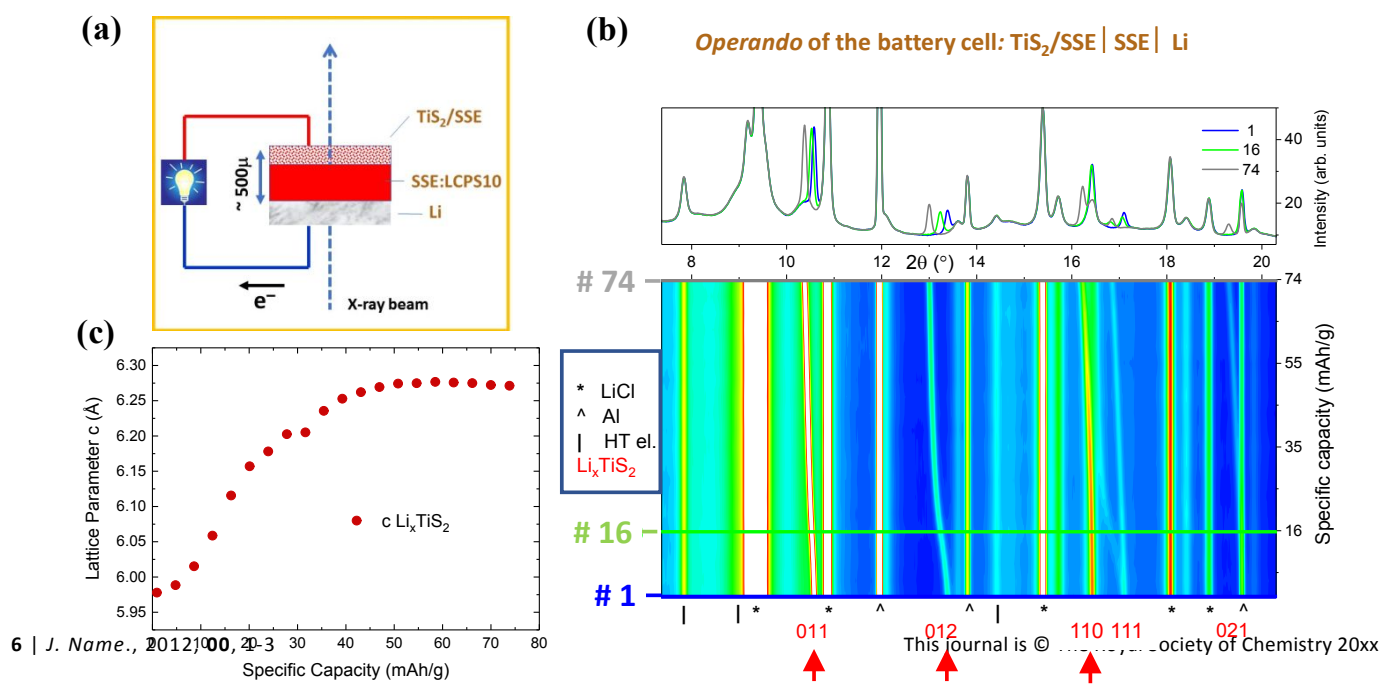

Figure 7. (a) Sketch of the experimental set-up during the operando data record. (b) 2D *operando* SR-PXD patterns of LCPS10-based battery cell during discharge (partial lithiation of TiS$_2$ electrode, C/10, 60 °C). Selected extracted 1D patterns at different cycling stages are shown on the top. (c) Evolution of the $c$-axis during lithiation of TiS$_2$. Al foil is used as current collector.

## Conflicts of interest

There are no conflicts to declare.

## Acknowledgements

This work is financially supported by Research Council of Norway under the program ENERGIX, Project no. 244054, LiMBAT. The authors also acknowledge the Research Council of Norway (Grant agreement SELiNaB-255441) for financial support. We acknowledge the skillful assistance from the staff of SNBL at ESRF, Grenoble, France. This work contributes to the research performed at CELEST (Center for Electrochemical Energy Storage Ulm-Karlsruhe).

## References

1. A. Mauger, C. M. Julien, A. Paolella, M. Armand and K. Zaghib, *Materials*, 2019, **12**, 3892.
2. M. Armand and J. M. Tarascon, *Nature*, 2008, **451**, 652-657.
3. J. B. Goodenough, *J. Solid State Electrochem.*, 2012, **16**, 2019-2029.
4. E. Quartarone and P. Mustarelli, *Chem. Soc. Rev.*, 2011, **40**, 2525-2540.
5. N. Legrand, B. Knosp, P. Desprez, F. Lapicque and S. Raël, *J. Power Sources*, 2014, **245**, 208-216.
6. C. Masquelier, *Nat. Mater.*, 2011, **10**, 649-650.
7. N. Kamaya, K. Homma, Y. Yamakawa, M. Hirayama, R. Kanno, M. Yonemura, T. Kamiyama, Y. Kato, S. Hama, K. Kawamoto and A. Mitsui, *Nat. Mater.*, 2011, **10**, 682-686.
8. M. H. Braga, N. S. Grundish, A. J. Murchison and J. B. Goodenough, *Energy Environ. Sci.*, 2017, **10**, 331-336.
9. A. Lecocq, G. G. Eshetu, S. Grugeon, N. Martin, S. Laruelle and G. Marlair, *J. Power Sources*, 2016, **316**, 197-206.
10. Y. Wang, W. D. Richards, S. P. Ong, L. J. Miara, J. C. Kim, Y. Mo and G. Ceder, *Nat. Mater.*, 2015, **14**, 1026-1031.
11. A. El kharbachi, Y. Hu, M. H. Sørby, J. P. Mæhlen, P. E. Vullum, H. Fjellvåg and B. C. Hauback, *Solid State Ionics*, 2018, **317**, 263-267.
12. B. Didier, N. Angeloclaudio, S. Dadi, E. T. M., V. M. H. W., Suwarno, V. Tejs, K. A. P. M. and d. J. P. E., *Adv. Funct. Mater.*, 2015, **25**, 184-192.
13. R. Mohtadi and S.-i. Orimo, *Nature Reviews Materials*, 2016, **2**, 16091.
14. P. E. de Jongh, D. Blanchard, M. Matsuo, T. J. Udovic and S. Orimo, *Appl. Phys. A*, 2016, **122**, 251.
15. M. Matsuo and S.-i. Orimo, *Adv. Energy Mater.*, 2011, **1**, 161-172.
16. Y. S. Choi, Y.-S. Lee, K. H. Oh and Y. W. Cho, *PCCP*, 2016, **18**, 22540-22547.
17. J. A. Teprovich, H. Colon-Mercado, A. L. Washington Ii, P. A. Ward, S. Greenway, D. M. Missimer, H. Hartman, J. Velten, J. H. Christian and R. Zidan, *Journal of Materials Chemistry A*, 2015, **3**, 22853-22859.
18. W. S. Tang, M. Matsuo, H. Wu, V. Stavila, W. Zhou, A. A. Talin, A. V. Soloninin, R. V. Skoryunov, O. A. Babanova, A. V. Skripov, A. Unemoto, S.-I. Orimo and T. J. Udovic, *Advanced Energy Materials*, 2016, **6**, n/a-n/a.
19. S. Kim, N. Toyama, H. Oguchi, T. Sato, S. Takagi, T. Ikeshoji and S.-i. Orimo, *Chem. Mater.*, 2018, **30**, 386-391.
20. S. Kim, H. Oguchi, N. Toyama, T. Sato, S. Takagi, T. Otomo, D. Arunkumar, N. Kuwata, J. Kawamura and S.-i. Orimo, *Nat. Commun.*, 2019, **10**, 1081.
21. A. Unemoto, H. Wu, T. J. Udovic, M. Matsuo, T. Ikeshoji and S.-i. Orimo, *Chem. Commun.*, 2016, **52**, 564-566.
22. A. Yamauchi, A. Sakuda, A. Hayashi and M. Tatsumisago, *J. Power Sources*, 2013, **244**, 707-710.
23. A. El kharbachi, Y. Hu, K. Yoshida, P. Vajeeston, S. Kim, M. H. Sørby, S.-i. Orimo, H. Fjellvåg and B. C. Hauback, *Electrochim. Acta*, 2018, **278**, 332-339.
24. J. P. Soulié, G. Renaudin, R. Cerný and K. Yvon, *J. Alloys Compd.*, 2002, **346**, 200-205.
25. Y. Filinchuk, D. Chernyshov and R. Cerny, *J. Phys. Chem. C*, 2008, **112**, 10579-10584.
26. A. El Kharbachi, E. Pinatel, I. Nuta and M. Baricco, *Calphad*, 2012, **39**, 80-90.
27. P. Vajeeston, P. Ravindran, A. Kjekshus and H. Fjellvåg, *J. Alloys. Compd.*, 2005, **387**, 97-104.
28. T. Ikeshoji, E. Tsuchida, K. Ikeda, M. Matsuo, H.-W. Li, Y. Kawazoe and S.-i. Orimo, *Appl. Phys. Lett.*, 2009, **95**, 221901.
29. M. Matsuo, Y. Nakamori, S.-i. Orimo, H. Maekawa and H. Takamura, *Appl. Phys. Lett.*, 2007, **91**, 224103.
30. T. Ikeshoji, E. Tsuchida, T. Morishita, K. Ikeda, M. Matsuo, Y. Kawazoe and S.-i. Orimo, *Phys. Rev. B*, 2011, **83**, 144301.
31. V. Epp and M. Wilkening, *Phys. Rev. B*, 2010, **82**, 020301.
32. H. Maekawa, M. Matsuo, H. Takamura, M. Ando, Y. Noda, T. Karahashi and S.-i. Orimo, *J. Am. Chem. Soc.*, 2009, **131**, 894-895.
33. R. Miyazaki, T. Karahashi, N. Kumatani, Y. Noda, M. Ando, H. Takamura, M. Matsuo, S. Orimo and H. Maekawa, *Solid State Ionics*, 2011, **192**, 143-147.
34. L. H. Rude, E. Groppo, L. M. Arnbjerg, D. B. Ravnsbæk, R. A. Malmkjær, Y. Filinchuk, M. Baricco, F. Besenbacher and T. R. Jensen, *J. Alloys Compd.*, 2011, **509**, 8299-8305.
35. L. H. Rude, O. Zavorotynska, L. M. Arnbjerg, D. B. Ravnsbæk, R. A. Malmkjær, H. Grove, B. C. Hauback, M. Baricco, Y. Filinchuk, F. Besenbacher and T. R. Jensen, *Int. J. Hydrogen Energy*, 2011, **36**, 15664-15672.
36. M. Matsuo, H. Takamura, H. Maekawa, H.-W. Li and S.-i. Orimo, *Appl. Phys. Lett.*, 2009, **94**, 084103.
37. L. M. Arnbjerg, D. B. Ravnsbæk, Y. Filinchuk, R. T. Vang, Y. Cerenius, F. Besenbacher, J.-E. Jørgensen, H. J. Jakobsen and T. R. Jensen, *Chem. Mater.*, 2009, **21**, 5772-5782.

38. K. Yoshida, S. Suzuki, J. Kawaji, A. Unemoto and S. Orimo, *Solid State Ionics*, 2016, **285**, 96-100.
39. D. Sveinbjörnsson, J. S. G. Myrdal, D. Blanchard, J. J. Bentzen, T. Hirata, M. B. Mogensen, P. Norby, S.-I. Orimo and T. Vegge, *J. Phys. Chem. C*, 2013, **117**, 3249-3257.
40. Z. Liu, M. Xiang, Y. Zhang, H. Shao, Y. Zhu, X. Guo, L. Li, H. Wang and W. Liu, *Phys. Chem. Chem. Phys.*, 2020, **22**, 4096-4105.
41. Y. Bouhadda, S. Djellab, M. Bououdina, N. Fenineche and Y. Boudouma, *J. Alloys Compd.*, 2012, **534**, 20-24.
42. H. Benzidi, M. Lakhal, A. Benyoussef, M. Hamedoun, M. Loulidi, A. El kenz and O. Mounkachi, *Int. J. Hydrogen Energy*, 2017, **42**, 19481-19486.
43. A. Unemoto, T. Ikeshoji, S. Yasaku, M. Matsuo, V. Stavila, T. J. Udovic and S.-i. Orimo, *Chem. Mater.*, 2015, **27**, 5407-5416.
44. A. El Kharbachi, H. Uesato, H. Kawai, S. Wenner, H. Miyaoka, M. H. Sorby, H. Fjellvag, T. Ichikawa and B. C. Hauback, *RSC Advances*, 2018, **8**, 23468-23474.
45. P. López-Aranguren, N. Berti, A. H. Dao, J. Zhang, F. Cuevas, M. Latroche and C. Jordy, *J. Power Sources*, 2017, **357**, 56-60.
46. L. Zeng, K. Kawahito, S. Ikeda, T. Ichikawa, H. Miyaoka and Y. Kojima, *Chem. Commun.*, 2015, **51**, 9773-9776.
47. S. Das, P. Ngene, P. Norby, T. Vegge, P. E. de Jongh and D. Blanchard, *J. Electrochem. Soc.*, 2016, **163**, A2029-A2034.
48. K. Kisu, S. Kim, H. Oguchi, N. Toyama and S.-i. Orimo, *J. Power Sources*, 2019, **436**, 226821.
49. P. M. Abdala, H. Mauroy and W. van Beek, *J. Appl. Crystallogr.*, 2014, **47**, 449-457.
50. V. Dyadkin, P. Pattison, V. Dmitriev and D. Chernyshov, *J. Synchrotron Rad.*, 2016, **23**, 825-829.
51. J. Sottmann, R. Homs-Regojo, D. S. Wragg, H. Fjellvåg, S. Margadonna and H. Emerich, *J. Appl. Crystallogr.*, 2016, **49**, 1972-1981.
52. U. Atsushi, C. ChunLin, W. Zhongchang, M. Motoaki, I. Tamio and O. Shin-ichi, *Nanotechnology*, 2015, **26**, 254001.
53. A. El kharbachi, Y. Hu, M. H. Sørby, P. E. Vullum, J. P. Mæhlen, H. Fjellvåg and B. C. Hauback, *J. Phys. Chem. C*, 2018, **122**, 8750-8759.
54. O. Zavorotynska, M. Corno, E. Pinatel, L. H. Rude, P. Ugliengo, T. R. Jensen and M. Baricco, *Crystals*, 2012, **2**, 144.
55. J. E. Olsen, M. H. Sørby and B. C. Hauback, *J. Alloys Compd.*, 2011, **509**, L228-L231.
56. J. E. Olsen, P. Karen, M. H. Sørby and B. C. Hauback, *J. Alloys Compd.*, 2014, **587**, 374-379.
57. V. Gulino, M. Brighi, E. M. Dematteis, F. Murgia, C. Nervi, R. Černý and M. Baricco, *Chem. Mater.*, 2019, **31**, 5133-5144.
58. A. Ha Dao, P. López-Aranguren, R. Černý, O. Guiader, J. Zhang, F. Cuevas, M. Latroche and C. Jordy, *Solid State Ionics*, 2019, **339**, 114987.
59. A. Hayashi, S. Hama, H. Morimoto, M. Tatsumisago and T. Minami, *J. Am. Ceram. Soc.*, 2001, **84**, 477-479.
60. S. Chen, K. Wen, J. Fan, Y. Bando and D. Golberg, *J. Mater. Chem. A*, 2018, **6**, 11631-11663.
61. A. El Kharbachi, O. Zavorotynska, M. Latroche, F. Cuevas, V. Yartys and M. Fichtner, *J. Alloys Compd.*, 2020, **817**, 153261.
62. K. B. Harvey and N. R. McQuaker, *Can. J. Chem.*, 1971, **49**, 3282–3286.
63. S. Gomes, H. Hagemann and K. Yvon, *J. Alloys Compd.*, 2002, **346**, 206-210.
64. H. Muramatsu, A. Hayashi, T. Ohtomo, S. Hama and M. Tatsumisago, *Solid State Ionics*, 2011, **182**, 116-119.
65. J. E. Trevey, C. R. Stoldt and S.-H. Lee, *J. Electrochem. Soc.*, 2011, **158**, A1282.
66. C. Julien and G.-A. Nazri, *Solid State Batteries: Materials Design and Optimization*, Springer US, 1994.
67. S. A. Pervez, M. A. Cambaz, V. Thangadurai and M. Fichtner, *ACS Applied Materials & Interfaces*, 2019, **11**, 22029-22050.
68. Z. Ding, J. Li, J. Li and C. An, *J. Electrochem. Soc.*, 2020, **167**, 070541.

# Pseudo-ternary LiBH$_4$·LiCl·P$_2$S$_5$ system as structurally disordered bulk electrolyte for all-solid-state lithium batteries

A. El Kharbachi,[a,b] J. Wind,[a,c] A. Ruud,[c] A.B. Høgset,[a] M.M. Nygård,[a] J. Zhang,[d] M.H. Sørby,[a] S. Kim,[e] F. Cuevas,[d] S. Orimo,[e,f] M. Fichtner,[b,g] M. Latroche,[d] H. Fjellvåg,[c] B.C. Hauback [a]

[a] Institute for Energy Technology (IFE), P.O. Box 40, NO-2027 Kjeller, Norway
[b] Helmholtz Institute Ulm (HIU) Electrochemical Energy Storage, Helmholtzstr. 11, 89081 Ulm, Germany
[c] Centre for Materials Science and Nanotechnology, University of Oslo, P.O. Box 1126, Blindern, NO-0318 Oslo, Norway
[d] Univ Paris Est Creteil, CNRS, ICMPE, UMR7182, F-94320, Thiais, France
[e] Institute for Materials Research, Tohoku University, Sendai 980-8577, Japan
[f] WPI-Advanced Institute for Materials Research, Tohoku University, Sendai 980-8577, Japan
[g] Institute of Nanotechnology, Karlsruhe Institute of Technology (KIT), P.O. Box 3640, 76021 Karlsruhe, Germany

**Experimental**

Lab-PXD data were obtained with a Bruker AXS D8 Advance diffractometer equipped with a Göbel mirror and a LynxEye 1D strip detector. The patterns were obtained in a Debye–Scherrer geometry using Cu Kα radiation (1.5418 Å) and rotating glass capillaries, filled and sealed under Ar atmosphere. Phase identification from the PXD data was performed using the DIFFRAC.SUITE EVA software with the PDF-4 database.

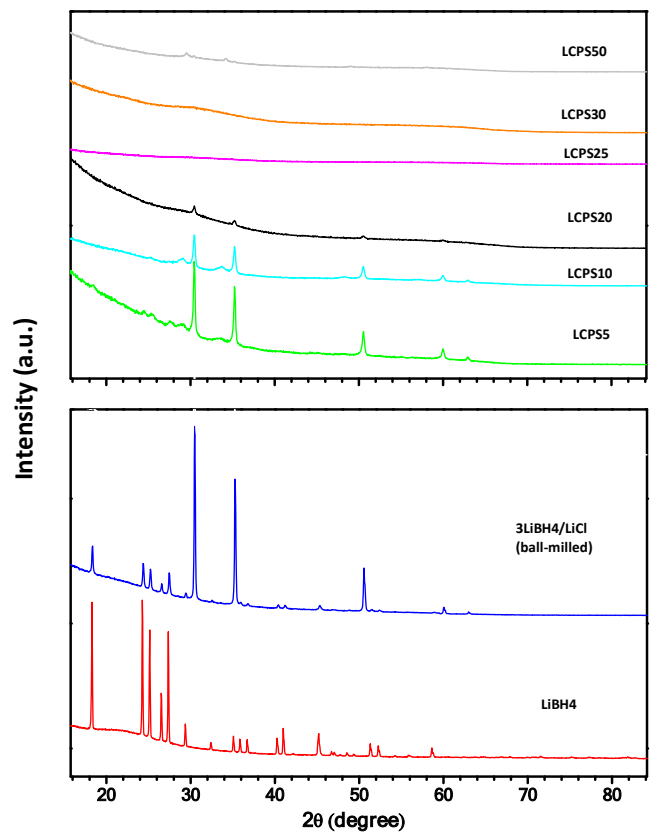

**Figure S1.** Lab-PXD patterns (Cu $K_\alpha$, 1.5418 Å) of LiBH$_4$, ball-milled 3LiBH$_4$-LiCl, and in the top figure LiBH$_4$-LiCl-P$_2$S$_5$ based mixed systems and further ball-milling with Spex milling; e.g. LCPS5 corresponds to 95(3LiBH$_4$/LiCl) - 5P$_2$S$_5$, and similar for the other compositions.

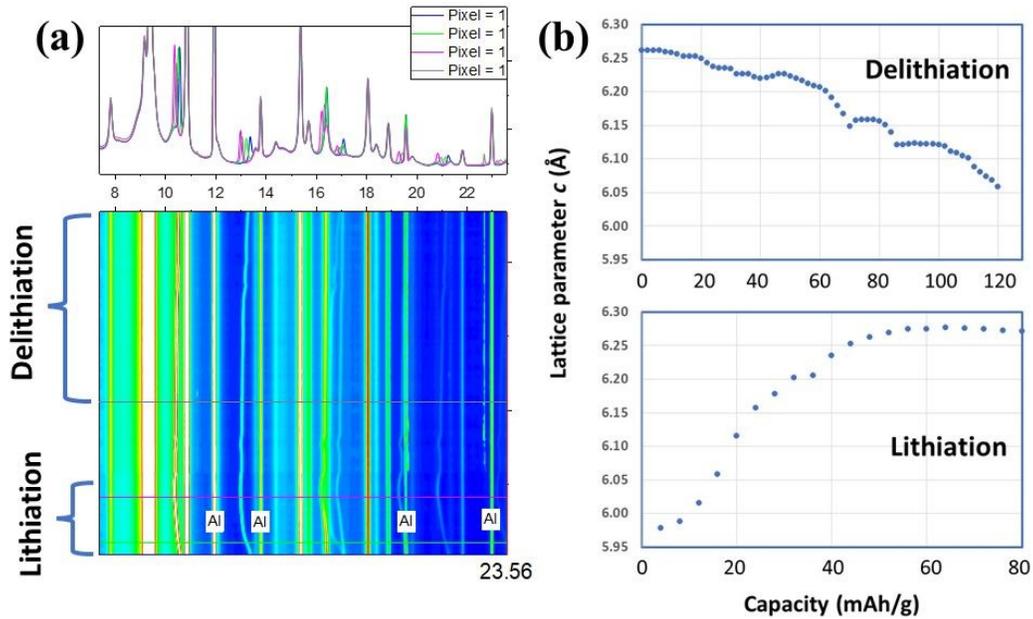

**Figure S2.** (a) 2D *in-operando* SR-PXD patterns of LCPS10-based battery cell during 1 cycle using TiS$_2$ electrode (extracted 1D pattern is shown on the top); (b) evolution of the *c*-axis during lithiation/delithiation of TiS$_2$. The cell was relaxed after lithiation for few hours while the PXD data are continuously collected.

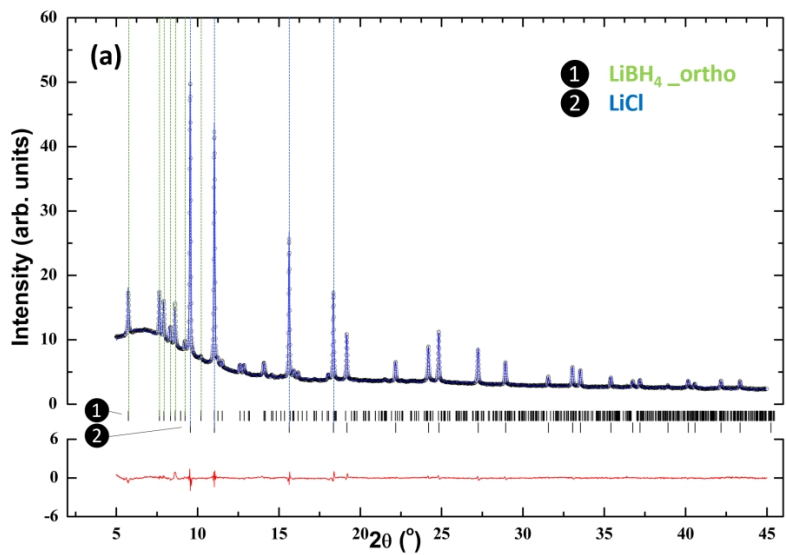

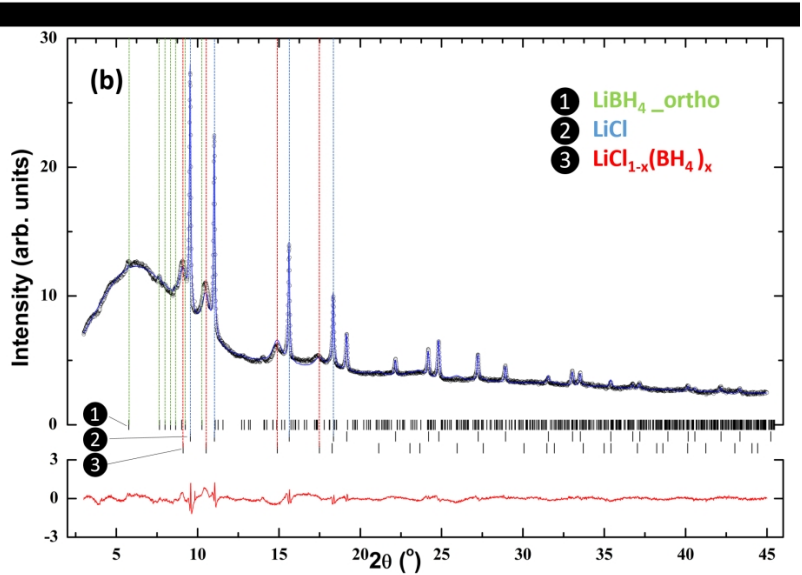

SR-PXD patterns of (a) 3LiBH4-LiCl as mixed after ball-milling and (b) LCPS10 solid electrolyte. The collected data at RT are shown in black, the fitted curve in blue, peak markers are shown in order for LiBH4, LiCl and LiBH4/LiCl, respectively. The difference plot (obs-calc) is shown in red at the bottom (λ = 0.4943 Å).

365x518mm (150 x 150 DPI)

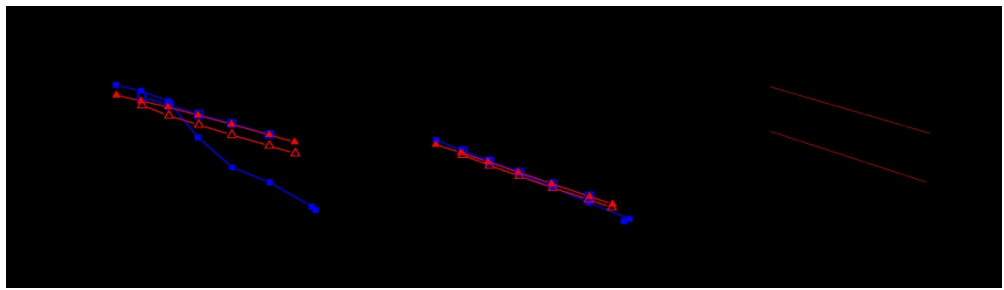

Temperature dependence of the ionic conductivities for (a) LCPS10 and (b) LCPS20 during the first two heating/cooling cycling T ramps. Filled/open squares and triangles correspond to heating/cooling for the 1st (blue) and 2nd (red) runs, respectively. (c) Arrhenius plots of the 2nd cooling used to infer the activation energies (Ea) for LCPS10 and LCPS20 samples.

551x156mm (94 x 94 DPI)

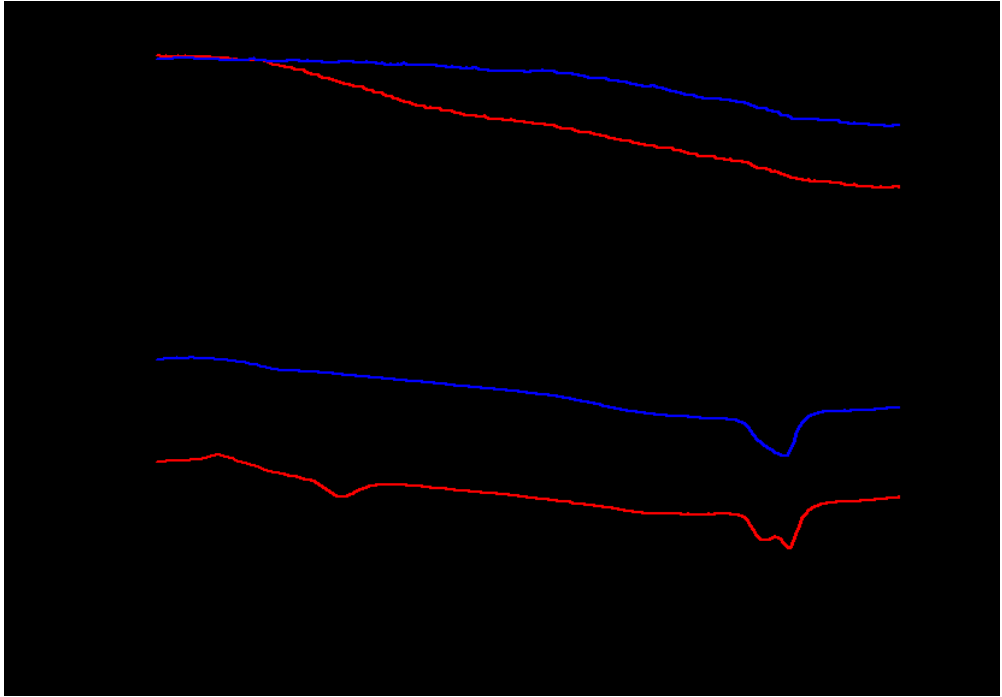

TGA (top) and DSC (bottom) analysis for (a) ball-milled LCPS10 and (b) annealed LCPS10 at 150 °C.

140x97mm (150 x 150 DPI)

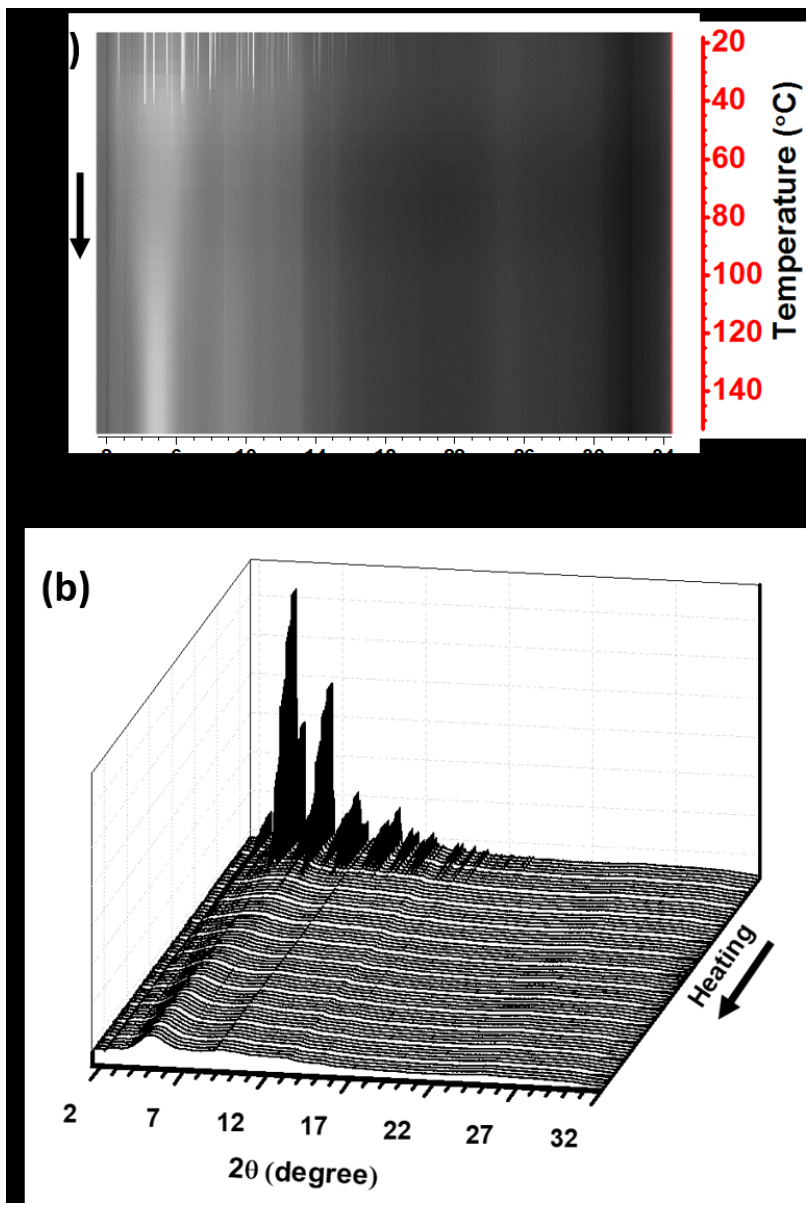

2D (a) and 3D (b) views of the SR-PXD of the LCPS10 as function of temperature from RT to 150 °C (10 °C min-1; λ =0.3123 Å). The linear heating scale is shown on the right Y-axis in (a).

130x194mm (150 x 150 DPI)

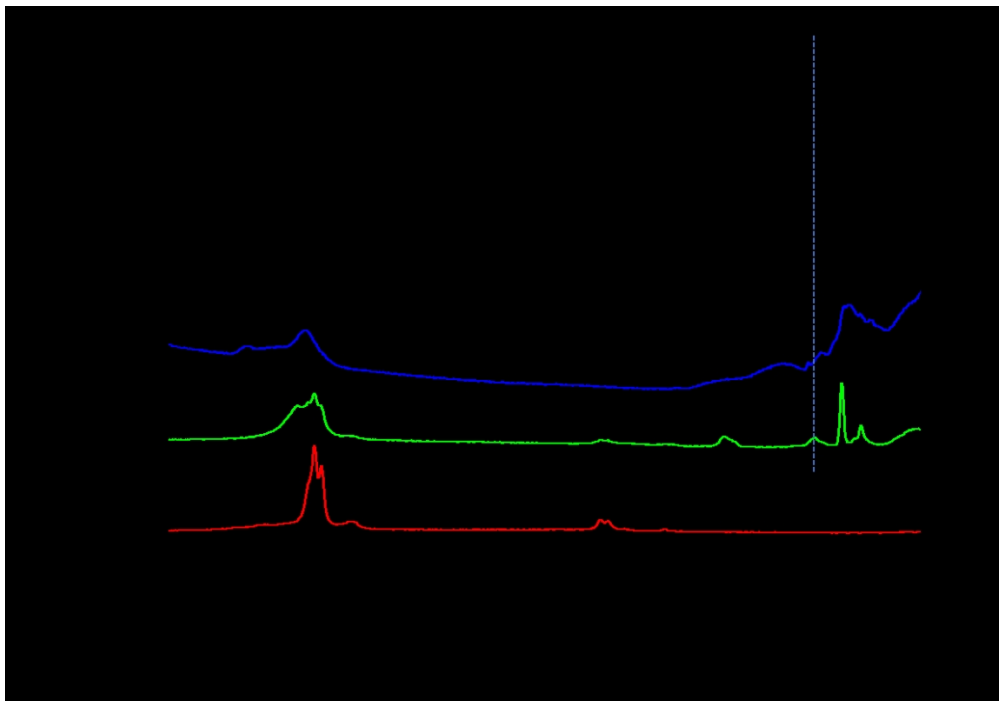

Raman spectra region 2800-210 cm-1 of (i) 3LiBH4-LiCl, (ii) LCPS10, (iii) LCPS20 and (iv) P2S5 samples.

199x139mm (150 x 150 DPI)

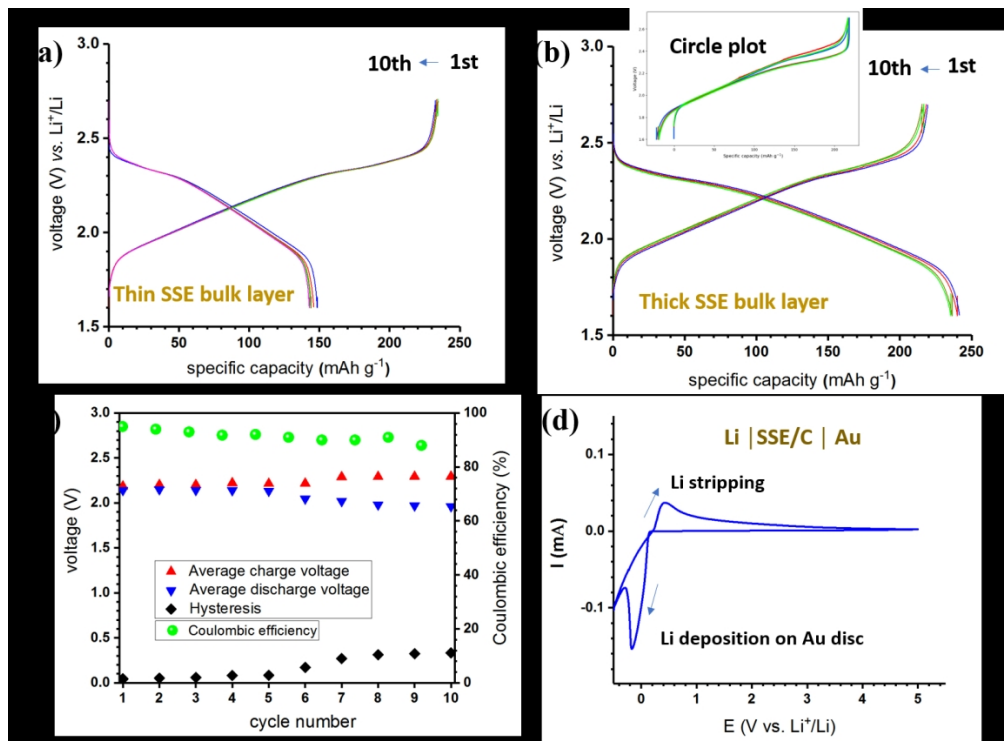

Electrochemical characterization and performance of the LCPS10 SSE: Galvanostatic discharge/charge cycling (0.05 C-rate) at 50 °C with TiS2 electrode (a) using thin (300 µm) and (b) thicker (600 µm) bulk layer of SSE; For the later cell, (c) average voltage, hysteresis (ΔV) and Coulombic efficiency as function of cycle number; (d) cyclic voltammetry at Au disc as working electrode vs. Li+/Li.

271x199mm (150 x 150 DPI)

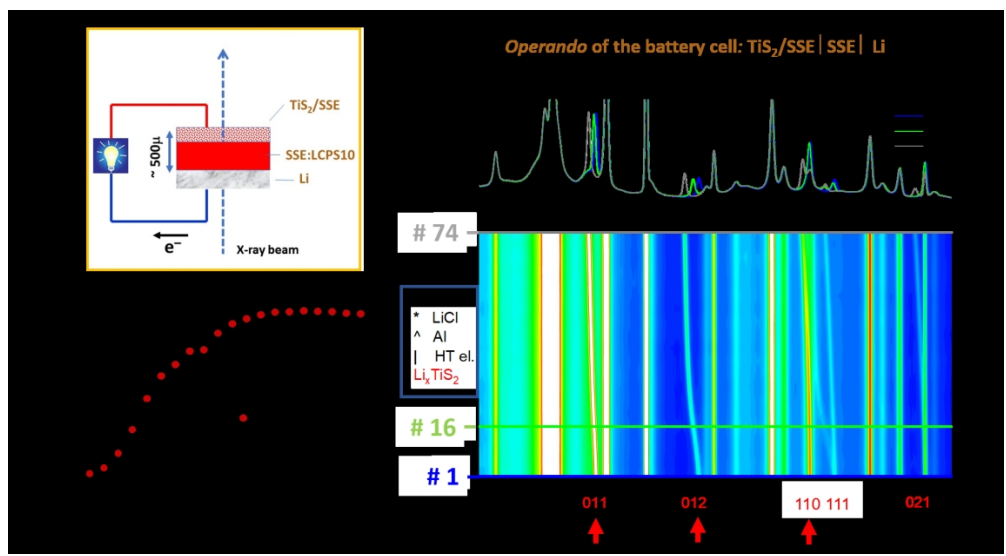

(a) Sketch of the experimental set-up during the operando data record. (b) 2D operando SR-PXD patterns of LCPS10-based battery cell during discharge (partial lithiation of TiS2 electrode, C/10, 60 °C). Selected extracted 1D patterns at different cycling stages are shown on the top. (c) Evolution of the c-axis during lithiation of TiS2. Al foil is used as current collector.

319x174mm (150 x 150 DPI)